\documentclass[useAMS,usenatbib]{mn2e}
\usepackage{times}
\usepackage{graphicx}
\usepackage{amssymb}

\title[The effect of local environment on dwarf galaxies]{Hubble Space Telescope survey of the Perseus Cluster -III: The effect of local environment on dwarf galaxies\thanks{Based on observations made with the NASA/ESA Hubble Space Telescope, obtained [from the Data Archive] at the Space Telescope Science Institute, which is operated by the Association of Universities for Research in Astronomy, Inc., under NASA contract NAS 5-26555. These observations are associated with programs GO-10201 and GO-10789}}
\author[S. J. Penny et al.]{Samantha~J.~Penny$^1$, Christopher~J.~Conselice$^1$, Sven~De~Rijcke$^2$, Enrico~V.~Held$^3$, \newauthor John S. Gallagher III$^4$ and Robert W. O'Connell$^{5}$   \footnotemark[0]\\ 
$^1$School of Physics \& Astronomy, University of Nottingham, Nottingham, NG7 2RD, United Kingdom\\
$^2$Sterrenkundig Observatorium, Universiteit Gent, Krijgslaan 281, S9, B-9000, Gent, Belgium\\
$^3$Osservatorio Astronomico di Padova, INAF, Vicolo Osservatorio 5, I-35122 Padova, Italy\\
$^4$Department of Astronomy, University of Wisconsin, 475 North Charter Street, Madison, WI 53706-1582, USA\\
$^5$ Astronomy Department, University of Virginia. P.O. Box 400325, Charlottesville, VA 22904-4325, USA}
\begin{document}

\maketitle

\begin{abstract}
We present the results of a \textit{Hubble Space Telescope (HST)} study of dwarf galaxies in the outer regions of the nearby rich Perseus Cluster, down to $M_{V} = -12$, and compare these with the dwarf population in the cluster core from our previous $HST$ imaging. In this paper we examine how properties such as the colour magnitude relation, structure and morphology are affected by environment for the lowest mass galaxies. Dwarf galaxies are excellent tracers of the effects of environment due to their low masses, allowing us to derive their environmentally based evolution, which is more subtle in more massive galaxies. We identify 11 dwarf elliptical (dE) and dwarf spheroidal (dSph) galaxies in the outer regions of Perseus, all of which are previously unstudied. We measure the ($V-I)_{0}$ colours of our newly discovered dEs, and find that these dwarfs lie on the same red sequence as those in the cluster core. The morphologies of these dwarfs are examined by quantifying their light distributions using CAS parameters, and we find that dEs in the cluster outskirts are on average more disturbed than those in the core, with $<$A$>$ $=0.13\pm0.09$ and $<$S$>$ $=0.18\pm0.08$, compared to $<$A$>$ $=0.02\pm0.04$, $<$S$>$ $=0.01\pm0.07$ for those in the core. Based on these results, we infer that these objects are ``transition dwarfs'', likely in the process of transforming from late-type to early type galaxies as they infall into the cluster, with their colours transforming before their structures. When we compare the number counts for both the core and outer regions of the cluster, we find that below $M_{V} = -12$, the counts in the outer regions of the cluster exceed those in the core. This is evidence that in the very dense region of the cluster, dwarfs are unable to survive unless they are sufficiently massive to prevent their disruption by the cluster potential and interactions with other galaxies. 
\end{abstract}

\section{Introduction}

Dwarf elliptical (dE) galaxies ($M_{\rm{B}} > -18$), and the fainter dwarf spheroidal (dSph) galaxies ($M_{\rm{B}} > -14$), are the most numerous galaxy type in the Universe. As such, any galaxy evolution/formation scenario must clearly be able to predict and describe the properties of these galaxies. As well as being numerous, dwarfs have low masses ($<10^{9}$ M$_{\odot}$), and are therefore more easily influenced by their environment than more massive galaxies, with environment likely playing a vital role in their formation and evolution. 

One well-known environmental dependence that dwarfs follow is the morphology-density relation defined by giant galaxies \citep{dressler80}, such that the fraction of early type galaxies increases with increasing environmental density. It was noted by \citet{reaves62} that dwarf irregulars were not found in great numbers in the Virgo Cluster, despite such galaxies being numerous in the Local Group, providing the first hints that an environmental dependence of dwarf galaxy morphology exists. The morphology-density relation was subsequently shown to extend to the dwarf galaxy regime in \citet{binggeli87}, with the number fraction of dwarf ellipticals increasing with local projected density. This relationship hints that the formation of dwarfs may be linked to the environment in which they reside. However, this environmental dependence of dwarf evolution is complicated by the fact that dwarfs in the cores of rich clusters are not a homogeneous population \citep{poggianti01,me08}, showing variations in their star formation histories.

We previously examined the star formation history of dwarfs in the core of Perseus in \citet{me08} using deep Keck/LRIS spectroscopy down to $M_{B} = -12$. These dwarfs exhibit a range of ages and metallicities, with an old, metal poor population with ages $\sim$8 Gyr, and a young, metal-rich population with ages $<5$ Gyr. This spread in ages and metallicities suggests multiple formation scenarios are required to explain their origin.

It is hypothesized that the spread in ages and metallicities can be explained if at least some dwarfs are the result of the morphological transformation of infalling galaxies. Several formation scenarios have been proposed to explain the infall origin of such cluster dwarf ellipticals, with a number of environmental processes thought to influence the formation and evolution of these galaxies. Dwarf galaxies are likely susceptible to high speed encounters with other galaxies (harassment, \citealt{moore}), ram pressure stripping by the hot ISM (e.g. Grebel et al. 2003), or by interactions with the tidal potential of their host cluster, group, or galaxy. These processes can cease star formation and remove mass from an infalling progenitor galaxy, transforming it from late to early type. The efficiency of these processes depend on the density of the environment in which the dwarf resides. However, the degree to which environment is important in shaping these low mass galaxies is largely unknown.

We began to investigate how environment shapes the evolution of dwarf galaxies in \citet{me09}. We investigated the dSph population in the core of the Perseus Cluster, using deep \textit{Hubble Space Telescope (HST)} Advanced Camera for Surveys (ACS) imaging in the F555W and F814W bands. This imaging reveals that dwarfs in the core of the Perseus Cluster are remarkably smooth in appearance, lacking the internal features such as spiral arms and bars seen in observations of brighter dwarf ellipticals ($M_{V} < -16$) in the Virgo and Fornax clusters \citep{jerjen00,sven03,Lisker06}. This result suggests these galaxies have not recently been transformed from late-type progenitors. Similarly, \citet{lisker09} find that dwarfs on circular orbits in the Virgo cluster are on average rounder than those on more elongated orbits, with no evidence of disks or residual star formation. Based on these results, they imply that these round dEs are part of the original cluster population, whereas the flatter dEs on elongated orbits originate from an infalling population. 

This study of how environment shapes low mass galaxies was continued in \citet{sven09}, where we compared the photometric scaling relations for early-type galaxies ($-8 \gtrsim M_{V} \gtrsim -24$) in different environments. This paper is based on our $HST$ ACS results, as well as data gathered from the literature, comprising of data for a range of environments from the Local Group to dense clusters. The results of this paper reveal that scaling relations for dSphs/dEs are almost independent of environment, however these scaling relations change at $M_{V} = -14$. This suggests that below this luminosity, the low masses, and therefore shallow potential wells, of these galaxies mean different processes likely drive their evolution. 

In this paper we continue our study of dwarf galaxies in the cluster environment, extending our study of the Perseus Cluster to its outer regions. Using deep \textit{Hubble Space Telescope (HST)} Wide Field Planetary Camera 2 (WFPC2) observations in the F555W and F814W bands, we target seven fields in the outer regions of the Perseus Cluster down to $M_{V}$ = -12. This imaging allows us to compare dwarfs in the core and outskirts of a rich galaxy cluster, to examine the effect of environmental density on the cluster dwarf galaxy population. Using this imaging, we identify eleven new dE/dSph systems in the outer regions of Perseus, all located at cluster-centric radii $> 300$ kpc. We also identify in this new imaging ten possible dIrr candidates, which we also discuss in this paper.

We examine the structures of our dEs and compare them to those of dEs in the cluster core. We find that the structures of dEs in Perseus become progressively more disturbed with increasing cluster-centric radius, which we interpret as evidence that these dwarfs are undergoing a morphological transition from late to early type galaxies as they infall into the cluster. However, despite being morphologically disturbed, these objects show no evidence for recent star formation, following the same colour magnitude relation (CMR) as those in the cluster core. This result suggests that these dwarfs have recently transformed from late-to-early type galaxies, with their colours transforming prior to their structures. We also compare number counts for both the core and outer regions of the cluster, and find that very faint dwarfs ($M_{V} < -12$) are preferentially found in the outer regions of Perseus.

This paper is organized as follows. In Section 2 we discuss our observations, including cluster membership in Section 2.5. The analysis of our data is presented in Section 3, with the morphologies of the dEs are covered in Section 3.1, the colour-magnitude relation as a function of environment discussed in Section 3.4, along with \textit{Galaxy Evolution Explorer (GALEX)} UV observations in Section 3.5. A comparison of the number counts in both the core and outskirts of the cluster are presented in Section 4, with a discussion of our results in Section 5. Our results are summarized in Section 6. Throughout this paper we refer to both cluster dEs and cluster dSphs as dEs.

\section{Data and Observations} 

\subsection{Target Cluster}

We examine the dwarf galaxy population of the Perseus Cluster (Abell 426), one of the nearest rich galaxy clusters, with a redshift $v = 5366$ kms$^{-1}$ \citep{Stublerood99}, and a distance $D = 72$ Mpc. Despite its proximity, it has not been studied in as much detail as other clusters such as Fornax, Virgo or Coma, most likely due to its low galactic latitude ($b$ = -13$^{\circ}$). 

To examine how local environmental density shapes the cluster dwarf galaxy population involves deep, high resolution imaging sampling the whole cluster, from its dense core to sparse outer regions. Using $HST$ ACS and WFPC2 observations spanning 800~kpc of the Perseus Cluster, we examine the effect of local environment on cluster dwarf ellipticals. Our survey design is shown in Fig. 1, with the ACS observations targeting the core of the cluster, and WFPC2 observations focusing on the cluster outskirts. The observations are explained in more detail below.

\subsection{ACS observations}

\begin{figure*}
\includegraphics[width=150mm]{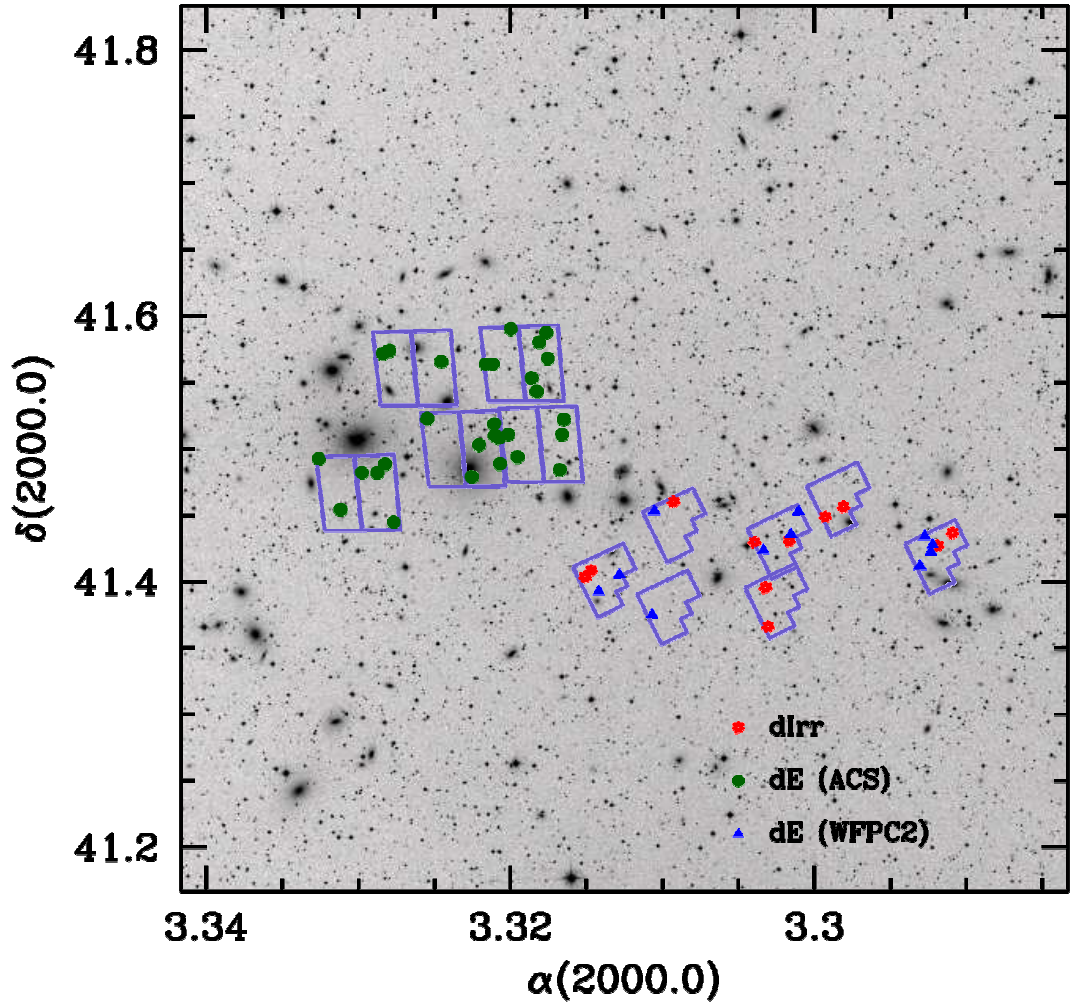}
\caption{Positions on the sky of the five $HST$/ACS (double rectangles) and seven $HST$/WFPC2 (chevrons) pointings, covering both the core and outer regions of the Perseus Cluster. Also marked are the postions of all dwarf ellipticals identified in this study. The fields and dE positions are overplotted on a DSS image spanning 600 kpc $\times$ 600 kpc of the cluster. }
\label{survey}
\end{figure*}

As part of the original study of dwarfs in the cluster core, we have obtained high-resolution $HST$ ACS Wide Field Camera imaging in the F555W and F814W bands for 5 fields in the core of the Perseus Cluster, resulting in a survey area of $\sim$57 arcmin$^{2}$ (program GO-10201). These pointings are in the vicinity of the peculiar galaxy NGC 1275 (taken to be central galaxy of Perseus) and the giant elliptical NGC 1272, the two brightest members of the cluster. Exposure times are one orbit each in the F555W and F814W bands. The reduction of these data are explained in detail in \citet{me09} and \citet{sven09}. The positions on the sky for these five pointings are shown in Fig.~\ref{survey}.

\subsection{WFPC2 observations} 

Due to the failure of ACS in 2007, we obtained $HST$ WFPC2 imaging for 7 fields in the outer regions of the Perseus Cluster in the F555W and F814W bands (program GO-10789). The total WFPC2 survey area is $\sim$40 arcmin$^{2}$, and extends to a distance of 600~kpc from the centre of the Perseus Cluster, which is taken to be the galaxy NGC 1275 ($\alpha = 03:19:48.1$, $\delta = +41:30:42$). Exposure times are one orbit per field in the F555W and F814W bands, resulting in a total exposure time of 14 orbits. These data were obtained in four exposures per field, to remove the effect of cosmic rays, and other defects, which could masquerade as features in the dwarfs. 

The WFPC2 images were processed using standard IRAF STSDAS techniques for the reduction of space telescope data, with a drizzle.scale parameter of 0.5. A drizzle.pixfrac scale of 0.9 was selected. This produced a final pixel scale of 0.05'', corresponding to 18 pc at the distance of the Perseus Cluster

Total magnitudes in the F555W for all dwarfs in our ACS and WFPC2 imaging are measured out to their Petrosian radii \citep{me09}. The colours of each dwarf are then measured within the central 1'' aperture. We correct these magnitudes for interstellar reddening adopting the colour excess $E(B-V) = 0.171$ of \citet{schlegel}. The magnitudes and colours for the dwarfs in our ACS imaging are then converted to the $UBVRI$ system using the transformations of \citet{sirianni05}, allowing their ($V-I)_{0}$ colours to be determined. The transformations to the $V$ and $I$ bands for the WFPC2 dwarfs are those of \citet{wfpc2cal} for the gain $= 7$ setting. 

\subsection{\textit{GALEX} observations}

To determine if any of the dwarfs in either the core or outer regions of Perseus contain recent star-formation, we utilise deep \textit{Galaxy Evolution Explorer (GALEX)} data in the far-ultraviolet (FUV) and near-ultraviolet (NUV) bands down to $NUV$ $\sim$24.5. When combined with structural information, star formation can provide an indicator of whether or not a dwarf is in the process of being transformed from late to early-type. If this is the case, it may retain a low level of star formation until the transformation is complete and all star-forming material is removed. 

We detect 3 dwarfs (CGW40, CGW45 and CGW46; \citealt{cgw03}) in the NUV, all of which are in the cluster core. These dwarfs are all non-detections in the FUV. The UV data are in AB magnitudes. We correct these magnitudes for interstellar reddening using the colour excess $E(B-V) = 0.171$ of \citet{schlegel}. The $NUV$ magnitudes and $(NUV-V)$ colours of these dwarfs are presented in Table 1. 

Unfortunately the $GALEX$ observations are not deep enough to detect the dwarfs in our WFPC2 observations. However, had any of the more luminous dwarfs in this region supported intense star formation, they would have shown up in the $GALEX$ imaging with anomalously blue colours. The three dwarfs that we did detect in the Perseus Cluster core can be compared to those of dwarfs in other clusters, to investigate if the $(NUV-V)$ colour-magnitude relation is consistent between clusters ($\S$~3.5), as is observed for optical colours (e.g. \citealt{sven09}). These dwarfs have NUV colours of normal inactive dEs. 

\begin{table}
\caption{Dwarf galaxies in our ACS imaging with $GALEX$ observations. The $NUV$ magnitudes and $(NUV-V)$ colours are extinction corrected.}
\begin{tabular}{lcccc}
\hline
&  $\alpha$  & $\delta$ & $NUV$ & $(NUV-V)$ \\
 Galaxy & (J2000.0) & (J2000.0) & (mag) & (mag)\\
\hline
CGW 40 & 03 19 31.7 & +41 31 21.3 & 22.23 $\pm$ 0.43 & 3.55 $\pm$ 0.47\\
CGW 45 & 03 19 41.7 & +41 29 17.0 & 21.72 $\pm$ 0.28 & 3.97 $\pm$ 0.32\\
CGW 46 & 03 19 42.3 & +41 34 16.6 & 22.68 $\pm$ 0.45 & 3.14 $\pm$ 0.48\\
\hline
\end{tabular}
\end{table} 

\subsection{Cluster Membership}

The most reliable method for confirming cluster memberships is by measuring radial velocities, though this is not easily done for low surface brightness objects such as cluster dwarf ellipticals, requiring long integration times. The high resolution and depth of our HST imaging allows for the separation of cluster dwarfs and background galaxies based on sizes, structures and morphologies (e.g. \citealt{me09}, \citealt{trent09}). We also have confirmed cluster membership for six of the dwarfs in our ACS imaging based on spectroscopic observations \citep{me08}. These spectroscopically confirmed cluster members all have similar morphologies, with smooth, extended light envelopes, little or no internal structure and a low surface brightness. All dwarfs in this study are well resolved, thus if a galaxy has the same morphology as a spectroscopically confirmed cluster member, we assume it is also a cluster member. Dwarf ellipticals are diffuse and symmetric in appearance, whereas background galaxies often contain spiral structures or tidal features. Dwarf ellipticals can also be separated from background galaxies as they have a low surface brightness, whereas background galaxies do not. This method was utilized in \citet{me09}, resulting in a sample of 29 dwarf ellipticals in the core.

\subsubsection{Dwarf Ellipticals}

\begin{figure}
\includegraphics[width=82mm]{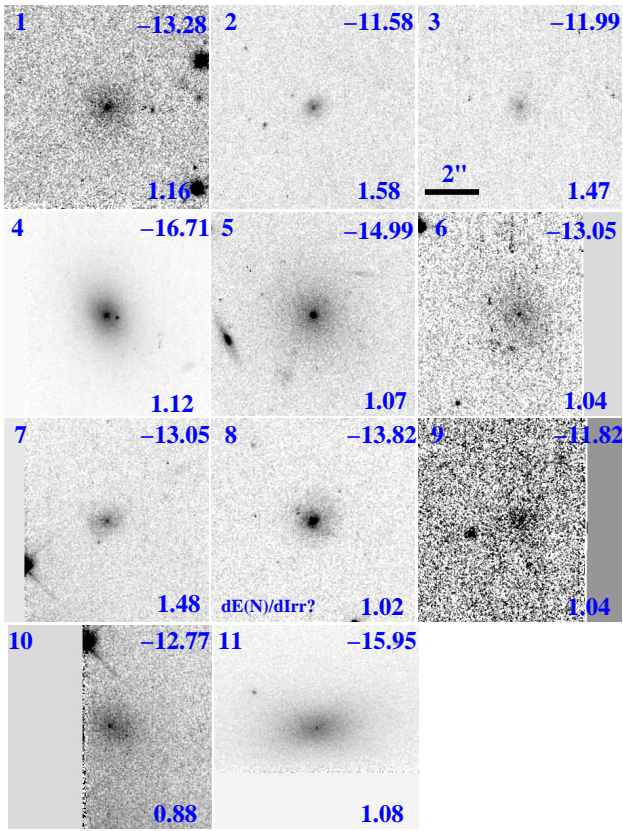}
\caption{Dwarf ellipticals in the outer regions of the Perseus Cluster. The upper right number is $M_V$, and the lower right-hand number is the $(V-I)_{0}$ colour. Dwarfs 2, 3 and 7 are very red, and we cannot rule out that these are background objects.}
\label{montage}
\end{figure}

We use the high resolution of our WFPC2 imaging to find dEs in a similar way to those in the cluster core \citep{me09}, identifying galaxies with similar morphologies to those with confirmed cluster membership in our ACS imaging. This yields a sample of 11 nucleated dwarf ellipticals (dE,N) in the cluster outskirts. The positions of these dwarfs within the Perseus Cluster are shown in Fig.~\ref{survey}, with a selection of images of these dwarfs presented in Fig.~\ref{montage}. A detailed assessment of these dwarfs, along with those in the cluster core, will be presented in Penny et al. (in prep). 

The presence of nuclei in all the galaxies in our cluster outskirts sample is consistent with our sample in the core, where 24 out of 29 ($\sim80\%$) dwarfs identified in the cluster core are nucleated down to $M_V = -12$. Previous surveys have suggested that below $M_{V} = -14$, nucleated dwarf ellipticals (dE,Ns) are rare, and more concentrated towards the cluster centre \citep{ferguson89}; however more recent studies have disputed this (e.g. \citealt{grant05}). The high fraction of faint nucleated dwarfs we have identified is comparable with recent space and ground based surveys of dEs (\citealt{grant05,cote06}). There is no evidence for this clustering in our dE sample, comparable to the findings of \citet{cote06}, who speculate that previous claims that dE,Ns are preferentially found in the cluster center may be due to selection effects, with only deep ground-based and space-based imaging able to resolve the nuclei of faint dEs \citep*{grant05,lotz05}. 

We further constrain cluster membership by quantifying the morphologies of all the dwarfs we identify by eye. Cluster dwarfs can be separated from background ellipticals based on quantitative measures of their structures, such as though their concentration indices \citep{cgw02,cas}. This index describes the ratio of the radii containing 80\% and 20\% of the galaxy's light, with the higher the value of C, the more centrally concentrated the light distribution is. Background ellipticals will have $C > 3.5$, whereas low surface brightness objects such as dwarfs will have C $\sim$2.5. All dwarfs we identify as dEs have $2.01 < C < 3.15$, showing they are low concentration galaxies rather than background ellipticals. However, one dwarf in the cluster core sample is found to have an unusually high concentration index (C$=5.385$), due to the presence of an unusually large nucleus, leading to ambiguity in its classification. This galaxy could therefore be a compact dwarf or a background galaxy. We therefore exclude this dwarf from our morphological analysis of the dE population of the cluster core and outskirts, though its colour places it on the red sequence of the Perseus Cluster colour magnitude relation. 

Size was also used to place constraints on cluster membership. Dwarf ellipticals have typical sizes $0.5$ kpc $ < r < 2.5$ kpc. Background spiral and elliptical galaxies at $z \approx 0.4$ are similar in angular size to dwarfs in Perseus (3.0''), though the high resolution of $HST$ imaging allows us to easily reject background spirals as cluster members. Likewise, background giant ellipticals will not have such a low surface brightness, will be more concentrated, and will be generally redder than cluster dEs.

\subsubsection{Dwarf Irregulars}

\begin{figure}
\includegraphics[width=84mm]{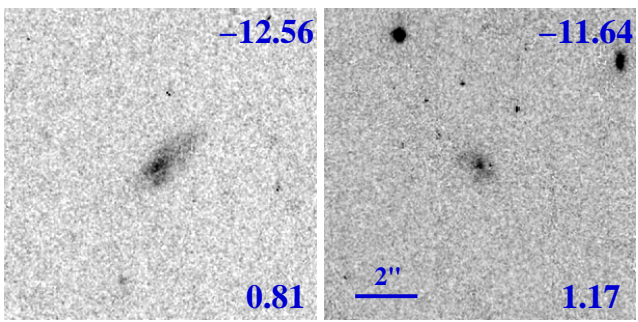}
\caption{Two example candidate dIrrs in the outskirts of the Perseus Cluster. The upper right hand number is the galaxy's total magnitude, and the lower right-hand number is its ($V-I$) colour. The candidate dIrrs do not exhibit the symmetric structures seen for cluster dEs. }
\label{dirrs}
\end{figure}

We identify in our WFPC2 imaging a sample of dwarf sized galaxies (0.5 kpc $< r <$ 2 kpc) with irregular morphologies that we cannot rule out as cluster members based on their colours, which are likely dIrrs. Two examples are shown in Fig.~\ref{dirrs}. However, their cluster membership will need to be confirmed via spectroscopy as these objects can in some cases appear similar to background spirals. We find no dIrrs in the ACS imaging of the cluster core. It is hypothesized that infalling lower mass late-type galaxies, such as dwarf irregulars (dIrrs), are the progenitors of dEs. Through tidal interactions as they infall into the cluster, these late-type galaxies lose their gas and cease star formation, and morphologically transform into early-type galaxies \citep{michielsen08,aguerri09,beasley09}. We identify dIrrs in our WFPC2 imaging based on their morphologies and colours. While we identify the dIrr candidates in Fig.~\ref{survey}, further results concerning the nature of these galaxies will be presented in Penny et al. (in preparation) once we can better assess the probability of their cluster membership. 

\section{Analysis} 

\subsection{Structures and morphologies}

It is hypothesized that some dwarf ellipticals are formed via tidal processes such as ram pressure stripping and the harassment of galaxies infalling into the cluster \citep{moore}. These transformations are thought to take place in environments with low velocity dispersions, such as galaxy groups which are later accreted by the cluster. The outskirts of a rich cluster represent a region where galaxies that have been recently accreted into the cluster are likely to reside. If the transformation from a late type galaxy to an early type dwarf elliptical does indeed take place in groups that are later accreted by the cluster, we might expect to find in the cluster outskirts a population of dwarfs that retain evidence of this transformation, such as remnant bars or tidal features (e.g. \citealt{sven02}).

Previous studies have found example objects, termed transition dwarfs (e.g. \citealt{grebel03}), which are galaxies that show characteristics of both early-type dwarfs (dEs \& dSphs) and late-types (dIrrs). It is likely that dIrrs are the precursors of dEs, with dEs being dIrrs that have lost their gas or converted it into stars long ago \citep{kormendy85,kormendy09}. Transition objects have little or no star formation and low luminosities (typical of dEs or dSphs), as well as containing HI gas as is typical for dIrrs. For example, \citet{bouchard07} find dwarfs in the Centaurus A Group with morphologies intermediate between early and late type that contain HI reservoirs. Similarly, \citet{cgw04} find that dwarfs in the Virgo Cluster with HI gas content are more similar to Local Group transition dwarfs than regular cluster dEs. Transition dwarfs likely represent a class of dwarfs undergoing the evolutionary transformation between late and early type galaxies, and may provide important clues as to the origin of dwarf ellipticals/spheroidals. These galaxies will have morphologies intermediate between dIrrs/dEs \citep{sandage91}, and we therefore examine the structures of the dwarfs identified in our WFPC2 imaging to identify such objects.

Morphological evidence for an infall origin will be erased over time as any internal sub-structure is removed (see $\S$~5). For infalling dwarfs to retain evidence of the transformation from late-type to early-type, they must have fallen into the cluster in recent times. By looking at the distances of our dwarfs from the cluster center, we can get an idea of the minimum time a dwarf must have been in the cluster for. A dwarf located at the edge of the cluster will either have just started to infall into the cluster, or be on an eccentric or highly radial orbit. This is because galaxies on circular orbits are likely part of the original cluster population, whereas those on radial orbits likely originate from an infalling population \citep{lisker09}. Combining this information with the morphologies of the dwarfs allows us to see if any of the dwarfs are being morphologically transformed as they infall into the cluster. For example, if these dwarfs are infalling into the cluster for the first time, and have disturbed morphologies, this would suggest they are in the process of transforming from late-type to early-type galaxies, with the cluster environment driving this transformation. 

The dwarfs in our WFPC2 imaging are all located at projected distances $>300$ kpc from the cluster centred galaxy NGC 1275, with the most distant dwarfs at distances $\sim$600 kpc from the cluster centre. If a morphologically disturbed dwarf is at a large distance from the cluster centre, it is either infalling into the cluster core for the first time, or else it is on a highly radial orbit. The typical crossing time for a dwarf entering a cluster such as Perseus will be $\sim$1.5 Gyr, therefore a dwarf at a distance of $\sim$600 kpc from the cluster centre is either entering the cluster for the first time, or has been in the cluster for $>$3 Gyr and is on a highly radial orbit, as it must have passed through its orbit pericenter already and returned to the cluster outskirts. Therefore the dwarfs in our WFPC2 imaging are either old objects or have been in the cluster for more than 3 Gyr. By quantifying their morphologies, we can see if the structures of the dwarfs in the cluster outskirts sample are consistent with a late-infall scenario.  

\subsection{CAS parameters}  

We quantify the morphologies of our newly identified dwarfs in the outer regions of Perseus using the concentration, asymmetry and clumpiness (CAS, \citealt{cas}) parameters. These parameters are measured using the inverted form of the Petrosian radius \citep{Pet76,Kron95,me09} for each galaxy. Each dwarf is manually cleaned of background galaxies and foreground stars using the {\sc{iraf}} task IMEDIT prior to the calculation of the CAS parameters, as these could masquerade as features in the dwarfs, or contaminate the measurements.

We compute the concentration index (C) for each galaxy, which is defined as the ratio of the radii containing 80$\%$ and 20$\%$ of a galaxy's light. The higher the value of C, the more concentrated the light of the galaxy is towards the centre, with dwarf ellipticals having an average value of 2.5 $\pm$ 0.3 \citep{cas}. The asymmetry indices (A) for the dwarfs in our sample are also computed. This index measures the deviation of the galaxies light from perfect 180$^{\circ}$ symmetry. The less symmetric the galaxy, the higher the asymmetry index. Asymmetric light distributions are produced by features such as star formation, galaxy interactions/mergers and dust lanes \citep{Conselice00}, which are not found in dwarf ellipticals in the core of Perseus \citep{me09}.

The third index we measure is the clumpiness index (S). This index describes how patchy the distribution of light within a galaxy is. The clumpiness (S) index is the ratio of the amount of light contained in high frequency structures to the total amount of light in the galaxy. To measure S, the original effective resolution of the galaxy is reduced by smoothing the galaxy by a filter of width $\sigma = R_{Petr}/5$ to create a new image. The original image is then subtracted by the new, lower resolution image to produce a residual map. The flux of the residual light is then summed, and this sum is divided by the sum of the flux in the original galaxy image to give the galaxy's clumpiness (S). For dwarf ellipticals in the core of Perseus, this is found to be near zero. However, if dwarf ellipticals in the outskirts of Perseus have only recently been transformed from late to early-type galaxies, then we might expect some internal substructure to remain. This is because young, blue stars have short life-times ($<100$ Myr, \citealt{bruzualcharlot03}) compared to the time-scale for star clusters and large scale star forming regions to dissolve ($> 1$ Gyr, e.g. \citealt{hasegawa08}). We give a detailed analysis of this situation in \S 5. This index is very sensitive to signal-to-noise, and we found it necessary to compute this index at half the Petrosian radius of each dwarf.

\begin{figure*}
\includegraphics[width=160mm]{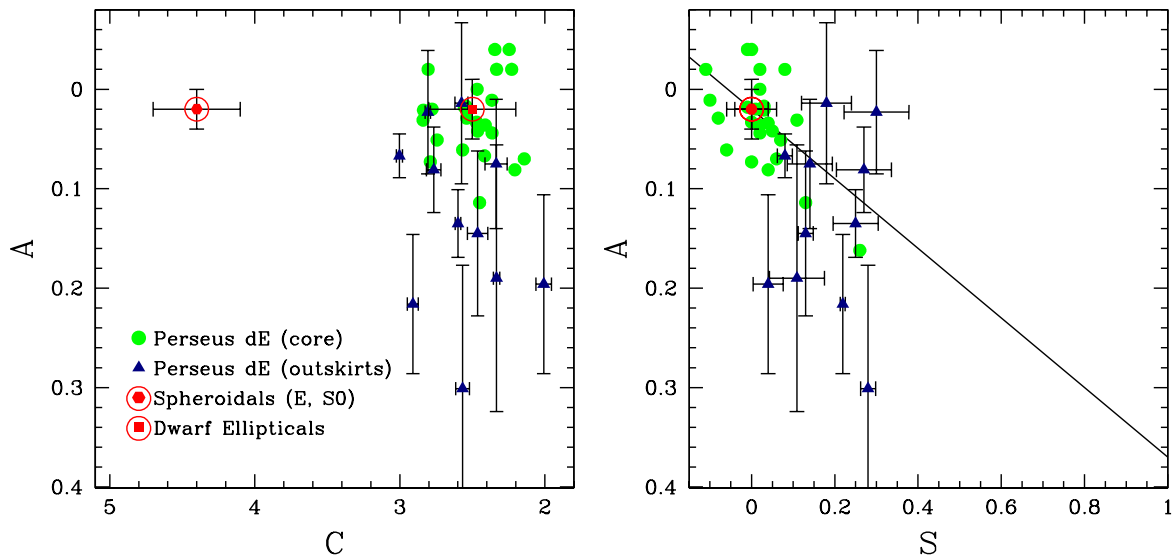}
\caption{Plots of asymmetry (A) versus concentration (C) and clumpiness (S) indices for dwarfs in the Perseus core (circles) and outskirts (triangles). Average error bars are provided. Also plotted are average C, A and S indices for dwarf ellipticals and spheroidals (ellipticals and S0s) from Conselice (2003) are plotted for comparison as circled points with error bars. The solid black line is the relationship between asymmetry and clumpiness from \citet{cas} The average values of both A and S are very similar, so do not appear at separate points in the plot of A vs S.}
\label{CAS}
\end{figure*}

We compare the CAS values for dwarfs in the core and outer regions in Fig.~\ref{CAS}. Dwarf ellipticals in the outskirts of Perseus have higher values of both A and S than those in the cluster centre, with dwarfs in the outer regions of the cluster having asymmetry and clumpiness values $<$A$>$ $=0.13\pm0.09$, and $<$S$>$ $=0.18\pm0.08$, compared to $<$A$>$ $=0.02\pm0.04$, $<$S$>$ $=0.01\pm0.07$ for those in the cluster core. These mean values were calculated using a bootstrap method. This method selects dwarfs at random for both the inner and outer regions of the cluster with replacement, drawing samples with the same size as the original data sets. This was repeated 10,000 times for both the inner and outer region data sets, giving bootstrap samples for the inner and outer regions of the cluster. We then compute a mean for each of the bootstrap resampled data sets, and calculate the dispersions for these means. The errors quoted are 1$\sigma$ from the mean. The dispersions for the means of these bootstrap samples are 0.024 and 0.010 for the outer and inner asymmetry means respectively, with dispersions for the outer and inner clumpiness bootstrap means of 0.024 and 0.013 respectively. These results indicate that dEs in the outer parts of Perseus are significantly more disturbed than those in the cluster core.

We also investigate the significance of this result using a Monte Carlo method. The Monte Carlo simulation we preform changes the values of the asymmetry and clumpiness values for dwarfs in the outer regions of the cluster by randomly varying the asymmetry and clumpiness values within the error bars for each galaxy. This is repeated 10,000 times, with the mean for each simulated outer data set compared to the mean for the inner dwarfs. What we find is that in none of these 10,000 realisations does the mean asymmetry of the simulated values for the outer dwarf asymmetries become lower than that of the inner dwarf galaxies.  In fact, we have to increase the sizes of the error bars by a factor of four before we see any increase in the fraction of the simulated data sets having an average asymmetry or clumpiness lower than those found for the inner sample.  For the clumpiness values, the error bars must be increased by a factor of ten. This implies the increase we see in the asymmetry and clumpiness values is robust to at least 10 $\sigma$ significance.

While the concentration index C does not reflect irregularities in a galaxy's light distribution, we compare the results for the two samples for completeness. We find that C is consistent between the two environments (c.f Fig.~\ref{CAS}), with $<$C$>$ $= 2.58 \pm 0.29$ for dwarfs in the cluster outskirts, compared to $<$C$>$ $= 2.52 \pm 0.21$ for those in the cluster core. These means were computed using the same method as for the asymmetry and clumpiness values. Therefore, within the error bars, the two populations have the same concentration values.

\subsection{Environmental Dependence of Morphology}

Local density decreases with increasing cluster-centric radius, such that the cores of clusters are more densely populated than the cluster outskirts. Galaxy morphology is dependent on local galaxy density, with spiral and irregular galaxies more commonly found in regions of low local density (the morphology-density relation, \citealt{dressler80,binggeli87}). Our HST survey areas are not large enough to accurately determine local galaxy density, so we instead use projected cluster-centric distance as an indicator for local galaxy density. 

\begin{figure*}
\includegraphics[width=160mm]{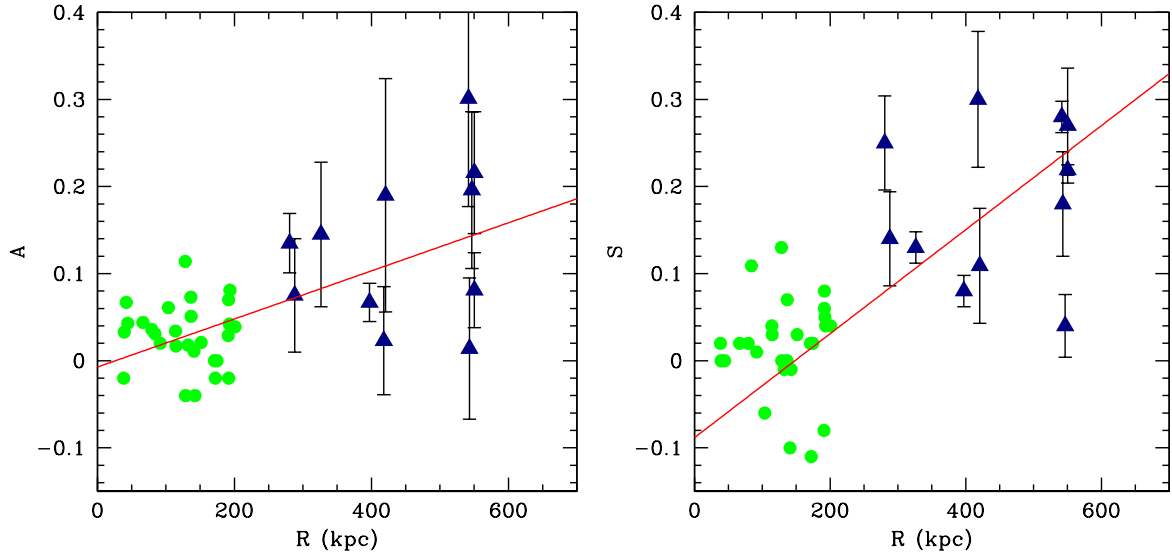}
\caption{Plots of A versus R (projected distance from the cluster center) and S versus R for the dwarfs in our sample. The circles are dwarfs in the core of the cluster, and the triangles are dwarfs in the outer regions. The red lines are least-squares fits to the data.}
\label{casr}
\end{figure*}

To determine how the morphologies of our dwarfs change with cluster-centric distance, we compute the projected distance of each dwarf from the cluster centre (taken to be the cluster centred galaxy NGC 1275), and compare these distances to their A and S values. The values of C are not investigated as a function of environment, as this index provides a useful way of separating cluster dEs from background ellipticals, but does not give a measure of features due to recent star formation or interactions \citep{cas}. The results are shown in Fig.~\ref{casr}. 

We fit the relationship between both A and S with R using a weighted least squares fit, using bootstrap samples. We generate 10,000 bootstrap samples, each of the same size as the original data set. A weighted least squares fit was then performed to each of the bootstrap samples. This process is repeated 10,000 times, with a new least-squares fit found for each resampling. We then find the mean and dispersions of these fits, with these best fit lines shown as red lines on Fig.~\ref{casr}. The errors quoted below are the dispersions of the least squares fits to the bootstrap samples. The following fits to the data are found:
\begin{equation}
A = (0.028\pm0.010)\left(\frac{R(kpc)}{100 kpc}\right) - (0.007\pm0.018),
\end{equation}
\begin{equation}
S = (0.060\pm0.024)\left(\frac{R(kpc)}{100 kpc}\right) - (0.088\pm0.053),
\end{equation}
where $R$ is the projected distance from the cluster centre. There is a correlation between both $A$ and $S$ with projected cluster centric distance $R$, such that the more distant a dwarf is from the cluster centre, the more disturbed its morphology is.

We confirm this result with a two-sided Kolmogorov-Smirnov test comparing the morphologies of dwarfs located at $R < 250$ kpc with those at $R > 250$ kpc from the cluster center. We compare the cumulative distributions of A and S for the core and outskirts environments. The likelihood that the values of A for the two environments are drawn from the same population is $0.33\%$, and $0.02\%$ for S. This low significance ($< 1\%$) shows that dwarfs in the outer parts of the cluster are indeed more disturbed than those in cluster core to a $2.5\sigma$ confidence level. However, not all dwarfs in the cluster outskirts have disturbed structures, with some dEs smooth like those in the core. This is to be expected as groups accreted into clusters will contain some smooth dwarfs (e.g. the Local Group dSphs). 

The slopes of the two relationships are similar, and this result suggests that the asymmetries in the dwarfs are the result of star formation rather than dynamical processes. The asymmetry and clumpiness indices are both good indicators of recent star formation, as star formation produces irregularities in the structure of the dwarf. The fact that the morphologies of the dwarfs become more disturbed with increasing cluster-centric radius can be explained by environmental effects. For example, ram pressure stripping varies as $\rho v^{2}$, where $\rho$ is the density of the intracluster medium and $v$ is the speed of the satellite galaxy relative to the ICM. Therefore this process is more effective towards the core of the cluster as the density of the intracluster medium increases. Galaxies at smaller clustercentric distances will have their gas removed more quickly than those in the cluster outskirts, so they cease star formation earlier. As star formation has ceased more recently for dEs in the cluster outskirts, their morphologies have not yet completely transformed from late to early type, unlike those at smaller clustercentric radii.

Alternatively, harassment \citep{moore96} can morphologically transform a small disk galaxy into a dwarf elliptical as it falls into the cluster. High-speed close encounters with massive cluster galaxies cause gravitational shocks that disrupt the disks of small Sc-Sd galaxies. The rate of these encounters will increase as the dwarf progenitor galaxy infalls into the cluster, as the density of the cluster (and therefore number of massive galaxies) increases. As a result, if a galaxy is transformed from late to early type via harassment, a galaxy that has just entered the cluster for the first time will have a more disturbed morphology than one in the core, as it will have had fewer encounters to destroy its internal substructure. Therefore a harassment formation scenario can also explain the more disturbed morphologies found for dwarfs in the cluster outskirts compared to those in the cluster core.  

\subsection{Colour magnitude relation}\label{CMR}

A well-defined colour-magnitude relation (CMR) exists for elliptical galaxies, such that the brightest galaxies are the reddest in colour, as first noted by \citet{sandage72} for ellipticals in the Virgo and Coma clusters. This correlation is generally regarded as a relationship between a galaxy's mass (as traced by luminosity), and metallicity (traced by colour, \citealt{smithcastelli,sven09}). Dwarf galaxies follow this same colour magnitude relation (e.g. \citealt{me08,sven09}). Using the newly identified dwarfs in the outer regions of Perseus, we investigate whether dwarfs in the lower density regions of rich clusters lie on this same colour-magnitude relation. Given that the morphologies of the dwarfs in the outskirts of Perseus suggest they have been recently transformed from late to early type, their colours may also reflect this, in that a recently transformed galaxy might be expected to be bluer in colour due to having a younger stellar population.

\begin{figure}
\includegraphics[width=82mm]{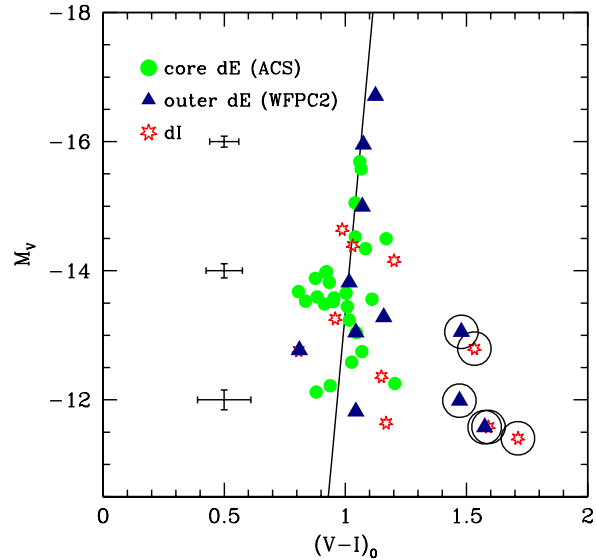}
\caption{Colour magnitude relation for dwarf ellipticals in Perseus. The circles represent dwarfs in the core of Perseus (ACS), the triangles represent those in the outer regions of the cluster (WFPC2), and the stars are dIrrs. The solid line is a linear fit to the colour magnitude relation for the dwarf elliptical population of Perseus as a whole, excluding any galaxies with ambiguous cluster membership (circled points, dwarfs 2, 3 and 7).}
\label{CMD}
\end{figure}

Our colour-magnitude relation is shown in Fig.~\ref{CMD}. We compare the CMR for the core and outer region dwarf populations, to examine any environmental dependence on the CMR. The CMRs for the two regions are fitted using a least squares method with the points weighted according to the error on their colour. This provides the following best fits to the data:

\noindent Core:
\begin{equation}
(V-I)_{0} = (0.76\pm0.30)-(0.017\pm0.022)M_{V},
\end{equation} 
\noindent and for the outskirts: 
\begin{equation}
(V-I)_{0} = (0.66\pm0.23)-(0.027\pm0.016)M_{V},
\end{equation}
Galaxies redder than $(V-I)_{0} = 1.3$ are excluded from this fit, due to the ambiguity of their cluster membership (circled points on the CMR, dwarfs 2,3 and 7). However, dwarfs this red have been found by \citet{mieske07} for fainter objects in the Fornax Cluster, which are more distant from the colour-magnitude relation than our candidate dwarfs, so they are included in our colour magnitude diagram. 

Within the error bars, the colour magnitude relations for the two environments are identical. This suggests that given the disturbed morphologies of the dwarf galaxy population the outskirts of Perseus, it appears that the colours of dwarfs transform prior to their morphologies. If this were not the case, we would expect the dwarf population in the outer regions to be significantly bluer than those in the core, with their colours matching better their disturbed morphologies. Since dwarf ellipticals in both the core and outer regions of the Perseus Cluster follow the same colour magnitude relation (Fig.~\ref{CMD}), this implies similar star formation histories and stellar populations (and therefore metallicities, \citealt{smithcastelli}) of dwarfs in the two environments. The dwarfs have colours consistent with passive stellar populations that stopped forming stars long enough ago for there to be no age information left \citep{smithcastelli,sven09}. 

Recent work by \citet{sven09} also suggests that all early-type galaxies follow a single CMR, regardless of environment, and we investigate if this relation extends to the outer regions of rich clusters. We perform a least squares fit to the colour magnitude relation for the cluster dwarf elliptical population as a whole (again excluding dEs redder than $(V-I)_0 = 1.3$), giving
\begin{equation}
(V-I)_{0} = (0.62\pm0.18)-(0.028\pm0.013)M_{V},
\end{equation}
The relationship between colour and absolute magnitude found by \citet{mieske07} for dwarfs in the Fornax Cluster core is $(V-I)_{0} = (0.52\pm0.07)-(0.033\pm0.004)M_{V}$, with a colour-magnitude relation of $(V-I)_{0} = 0.45 - 0.039M_{V}$ found for the Hydra I Cluster \citep{misgeld08}. A similar relation of $(V-I)_{0} = (0.33\pm0.02)-(0.042\pm0.001)M_{V}$ is found for the Centaurus Cluster. Within the error bars, the three clusters have identical colour-magnitude relations as Perseus for their dwarf galaxy populations. This also matches the colour magnitude relation found for dwarfs in the Local Group \citep{grebel03,sven09}. Therefore it appears that the universality of the CMR extends to dEs in the low-density, infall regions of clusters. 

We also include for comparison our dIrr candidates. Like the cluster dEs, all these galaxies are red in colour, and we can reject any candidates with ($V-I)_{0} > 1.6$ as these are probable background galaxies. The red colours of these galaxies suggests they are either background galaxies or dIrrs that have entered the cluster at recent times, and have been stripped of their gas and ceased star formation.

\subsection{\textit{GALEX} UV Observations}

We extend our study of the cluster colour-magnitude relation to the UV, by examining the $(NUV-V)$ colours of our dwarfs using GALEX data for the Perseus Cluster. If a dwarf has been recently transformed from late-to-early type, its stellar population will likely retain evidence of recent or ongoing star formation. During ram presssure stripping, the high pressure of the hot intracluster medium (ICM) can trigger the collapse of molecular clouds in an infalling late-type galaxy, leading to a burst of star formation \citep{bekki}. Likewise, tidal perturbations in the dense cluster core can trigger star formation episodes. An indication of whether recent star formation has taken place in our early-type dwarfs can therefore be gained by examining their ($NUV-V$) colours.

However, a flux in UV does not necessarily identify a young stellar population. Evolved stars, such as old, hot horizontal branch and post asymptotic giant branch stars with high metallicity \citep{bressan94} can produce a UV excess in early-type galaxies, and it is important to determine if such a population is producing the UV flux in the dwarfs. If a dwarf has undergone a recent burst of star formation, its (NUV$-$V) colour will reflect this, being bluer than the colour magnitude relation for early type galaxies.

\subsubsection{The ($NUV-V$) colour magnitude relation}

\begin{figure}
\includegraphics[width=82mm]{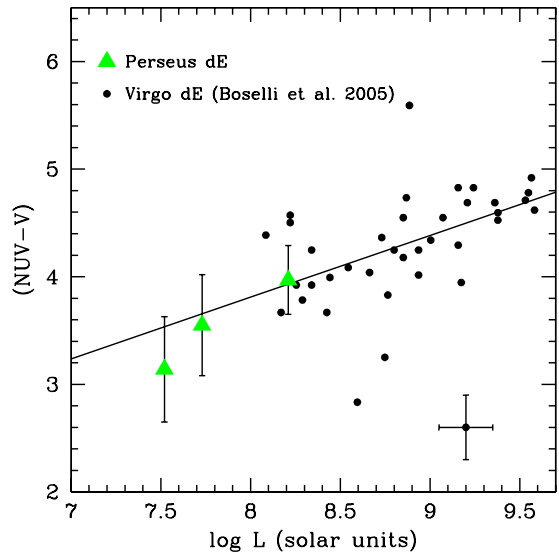}
\caption{($NUV-V$) colour magnitude relation for dwarfs in the Perseus Cluster (triangles), overplotted on the results of \citet{boselli} for dEs in the Virgo cluster. The Perseus dwarfs for which we have $NUV$ data are CGW 40, CGW 45 and CGW 46. The line is the CMR for the Virgo Cluster dEs, which we extend to fainter magnitudes.}
\label{UV}
\end{figure} 

\citet{boselli} investigate the ($NUV-V$) colour-magnitude relation for dwarf ellipticals in the Virgo Cluster. By comparing the colour-magnitude relation for the core of Perseus and for Virgo, we can see if the ($NUV-V$) CMR is consistent between clusters. To a limiting magnitude of M$_{V} = -15$, the ($NUV-V$) colour index for Virgo dwarfs reddens monotonically with increasing luminosity. We extrapolate this colour magnitude relation to $L_{V} = 10^{7.5}$ L$_{\odot}$, to cover the faintest dwarf (CGW46) for which we have NUV data, with the results presented in Fig.~\ref{UV}. The three dwarfs in the Perseus core lie on the same CMR as those in the Virgo cluster, and are not excessively blue, suggesting they have not undergone a recent burst of star formation.

Given that these dwarfs have remarkably smooth morphologies \citep{me09}, and that their stellar populations show no evidence for formation via a recent infall, it is unlikely that the UV detections of these objects are due to recent star formation.  Instead, this result suggests that the UV detections of these dwarfs is due to the presence of an old stellar population. This fact can be used to place a limit on the maximum star formation rate these detected dwarfs can have. The limiting magnitude of our $GALEX$ data is NUV = 24.5 ($M_{NUV} = -10$), and given that most of our dwarfs are non-detections in the UV, or they are detected due to the presence of an old, evolved stellar population, the UV flux due to star formation must therefore be fainter than this. The NUV luminosity of an object can be used to calculate a star formation rate, using the method of \citet{kennicutt98}. This gives a maximum star formation rate of $\sim0.002$ M$_{\odot}$yr$^{-1}$ for the dwarfs i.e. they must be forming fewer stars than this per year to be non-detections.

\section{Number Counts}

Due to their low masses, dwarf galaxies are thought to be particularly susceptible to tidal forces, ram pressure stripping and harassment. In the centre of the cluster, the cluster tidal potential becomes stronger, so it is expected that smaller spheroidal galaxies will be disrupted in this region. In the outer parts of the cluster, these tidal forces are weaker and dwarfs are therefore more able to survive the cluster potential, resulting in higher number counts and a steeper faint end to the luminosity function (e.g. \citealt{thompson,phillips98a,Pra,sanchezjanssen05}). However, in \cite{me09} we argue that most dwarfs currently in the cores of rich clusters must have a large dark matter component to prevent such disruption.

We compare the relative numbers of low mass galaxies in the core and outskirts of Perseus using number counts. All sources in our ACS and WFPC2 imaging were identified using SE\textsc{xtractor} \citep{SExtractor}. Objects with more than 50 contiguous pixels 2$\sigma$ above the sky rms were identified. We exclude from our ACS catalog all objects with central coordinates less than 50 pixels from the CCD edges and gaps, to prevent the inclusion of incomplete objects in the catalog. Edge objects were likewise excluded from the WFPC2 catalog.  

\subsection{Star-Galaxy Separation}

To provide accurate number counts, stars must be removed from our object catalogs. As the Perseus Cluster is at a low galactic latitude ($b = -13^{\circ}$), the object counts will be contaminated by large numbers of foreground stars. At the distance of Perseus, a dwarf galaxy with M$_{I} = -10$ will have an apparent magnitude $I = 24.5$, so we therefore require a reliable method to separate stars and galaxies down to this faint magnitude.

\begin{figure*}
\includegraphics[width=160mm]{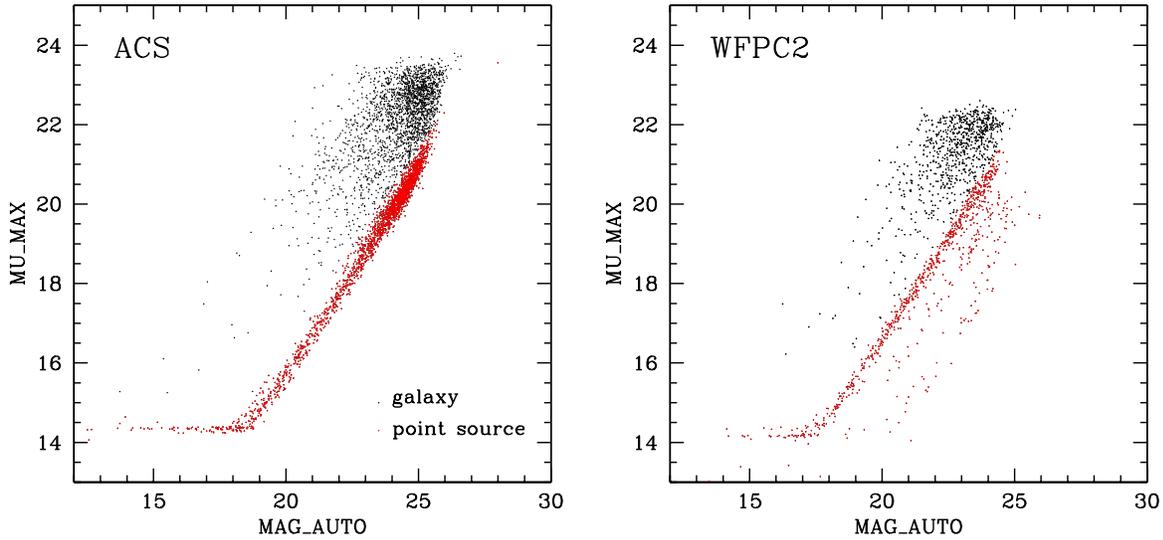}
\caption{Separation of stars (red points) and galaxies (black points) using \textsc{mu\_max} vs \textsc{mag\_auto}. Point sources follow the PSF, whereas extended sources (galaxies) lie above this relation. Point sources become indistinguishable from galaxies below F814W = 25.}
\label{sep}
\end{figure*}

One way to separate stars and galaxies is to use the SE\textsc{xtractor} stellar classification index \textsc{class\_star}, where a value of 0 corresponds to an extended source, and a value of 1 to a point source. However, for fainter sources the reliability of this classification breaks down, making the separation of stars and galaxies difficult for sources fainter than $I = 22$ (e.g. \citealt{cosmos}). 

The separation of galaxies and stars is instead done using the relationship between the SE\textsc{xtractor} parameters \textsc{mum\_max} and \textsc{mag\_auto}, where \textsc{mag\_auto} is the total magnitude, and \textsc{mu\_max} is the peak surface brightness of an object against the background level. This method was utilized by \citet{cosmos} to separate stars and galaxies in the COSMOS survey, and is based on the fact that the light distribution of point sources scales with magnitude. Point sources such as stars occupy a well defined region in the \textsc{mum\_max} vs \textsc{mag\_auto} plane as they follow the point spread function (PSF), as shown in Fig.~\ref{sep}, allowing us to separate stars and galaxies down to F814W $\sim 25$. This method does not distinguish between dEs and dIrrs, so both are included in our number counts.

\subsection{Number Counts}

\begin{figure*}
\includegraphics[width=164mm]{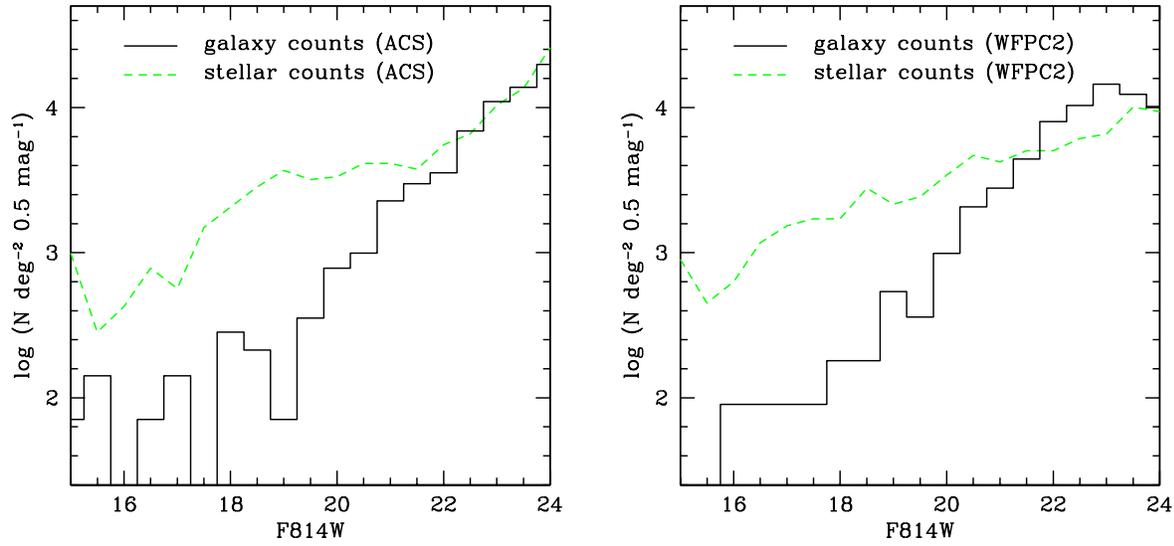}
\caption{Raw number counts (no background galaxy subtraction) for galaxies in both the ACS and WFPC2 survey areas. The dashed lines are the stellar number counts, and the histogram is the galaxy counts. These counts are normalized to an area of 1 deg$^{2}$ to allow comparison between the two surveys.}
\label{ncounts}
\end{figure*}

The galaxies are binned according to magnitude, with the results shown in Fig.~\ref{ncounts} for both the core and outer regions of the cluster, along with the stellar counts for both regions. The higher stellar counts in the ACS data at faint magnitudes due to the presence of large numbers of globular clusters around the giant ellipticals, with these globular clusters contributing to the point source counts. 

To measure an accurate faint-end slope to the galaxy luminosity function, it is essential to ensure that number counts are not contaminated by background galaxies. This can be done using spectroscopy to measure the redshift of the galaxies in question, or via the statistical subtraction of background galaxies. However, given that the same background counts should apply for both the core and outer regions, we can make a direct comparison for both the core and outer regions without making a background subtraction.

\begin{figure}
\includegraphics[width=82mm]{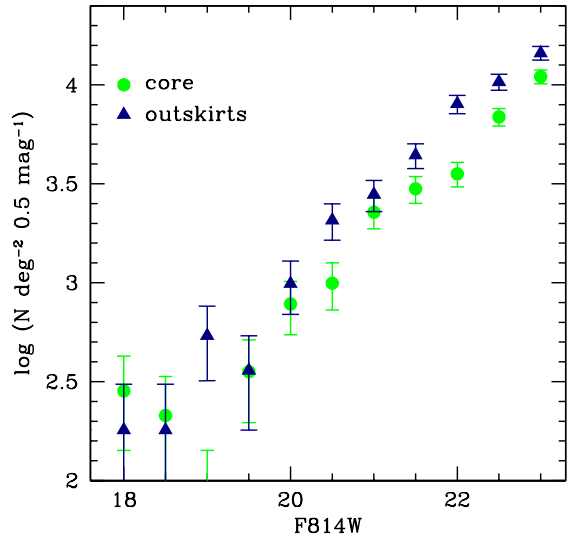}
\caption{Comparison of the number counts for objects in the core and outskirts of the Perseus Cluster. These counts are normalized to an area of 1 deg$^{2}$ to allow comparison between the two surveys.}
\label{ncountcomp}
\end{figure}

We compare the number counts for the core and outer regions in Fig.~\ref{ncountcomp}. The error bars are the standard deviation for each bin. Number counts below F814W = 23 are excluded as these are fainter than the completeness limits for the WFPC2 imaging. Within our error bars, the number counts for the two populations are identical for dwarfs brighter than $M_{I} = 21$. At F814W $>$ 21.5, the number counts in the outskirts of Perseus exceed those in the cluster core, suggesting that for at least the very faintest galaxies ($M_{I} < -13$, $\sim$$M$$_{V} = -12$) environment plays an important role in their evolution. One such role is that some dwarf ellipticals are unable to survive in the cores of rich clusters unless they are sufficiently dark matter dominated to prevent their disruption \citep{me09}. Dwarfs in the cluster outskirts are subject to smaller direct environmental influences, therefore low mass galaxies are more able to survive in this region. This results in a steeper faint-end slope to the luminosity function in lower density environments e.g. the `dwarf galaxy-density relation' suggested by \citet{phillips98}.

\section{Discussion}

We compare the morphologies and structures of dEs in both the core and outskirts of Perseus, and find those in the cluster outskirts to be on average more disturbed than those in the cluster core. This can be interpreted in two ways: either the dwarfs are systems undergoing the transformation from late-to-early type galaxies as they infall into the cluster, or they are on extremely radial orbits with small pericentric distances, resulting in them being morphologically disturbed via high speed interactions as they pass through the cluster core. Such high speed interactions are unlikely to destroy these galaxies unless they pass very close to the cluster core \citep{cgw01}, but are likely to significantly modify their structures. However, if the structures of such galaxies were indeed modified as they pass through their orbit pericenter, we would expect to see morphologically disturbed dwarfs in the cluster core, whereas in \citet{me09} we identify a remarkably smooth population of dEs in this region, with no evidence of such morphological disruption.

However, not all the dwarfs in the outskirts have disturbed morphologies- some dEs are smooth as are all of those in the core, with A $< 0.1$ and S $< 0.1$. It is thought that clusters grow via the accretion of galaxy groups (e.g. \citealt{cortese}), and groups such as the Local Group contain a fraction of dEs and dSphs, as well as dIrrs. Therefore at least some of the dwarfs entering the cluster will have smooth morphologies, and only those transforming from dIrrs or spirals will have disturbed morphologies. 

The dEs we identify lie on the colour magnitude relation expected for dEs in all environments, suggesting that if they do originate from an infalling population of galaxies transforming from late-type to early-type, they have lost their gas and therefore ceased star formation. We infer that these galaxies in the outer regions of Perseus are therefore ``transition dwarfs'', with the cluster environment likely playing an important role in modifying these galaxies (e.g. \citealt{gallagherhunter}). Other examples of such dwarfs have been found in the Virgo and Fornax clusters \citep{sven03,lisker09a}. These dwarfs lie on the red sequence, but do not have the smooth structures found for the dwarf ellipticals in the core of Perseus. This can be explained due to the timescales required for the colour and morphological transformation of a late-type to early-type galaxy. The process of removing gas from a star forming dwarf can take place over short timescales given the shallow potential wells of these galaxies. For example, ram-pressure stripping can remove the gas reservoir, shutting off star formation in a dwarf galaxy entering a cluster over a timescale of $\sim$100 Myr \citep{mori00,michielsen04}. Evidence for this gas removal has been seen, with the majority of dEs in the Virgo Cluster do not contain HI gas \citep{cgw04}, suggesting that the process of gas removal is very efficient in low mass galaxies, causing star formation to halt rapidly.

\begin{figure}
\includegraphics[width=82mm]{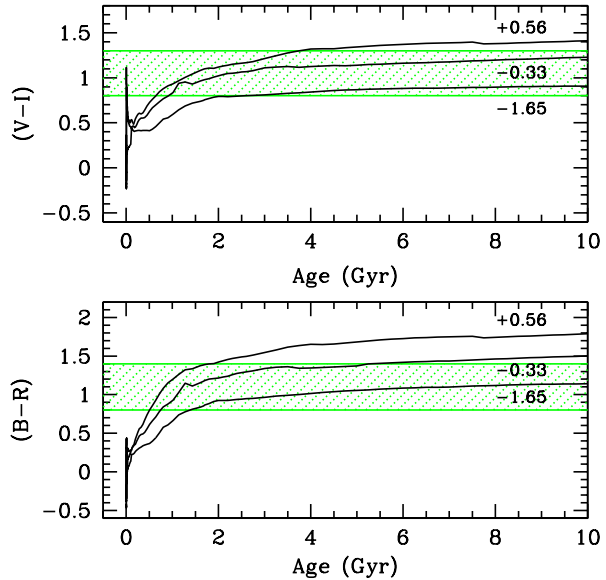}
\caption{The evolution of the \citet{bruzualcharlot03} models of the ($V-I$) and ($B-R$) colour indices with time. We include models for [Fe/H]$ = -1.65, -0.33,$ and $+0.56$. The shaded region represents the range in colours for our dwarfs in both environments, excluding the very red cluster candidates in our WFPC2 imaging due to the uncertainty in their cluster membership.}
\label{bruzualcharlot}
\end{figure}

The colours of star forming galaxies, and star forming regions within these galaxies, are dominated by young, blue O and B type stars. However, once their gas has been removed via ram pressure stripping or some other method, the reservoir of gas required to fuel this star formation no longer exists. Therefore the colour of the galaxy will rapidly redden due to the short main-sequence life-times of massive stars. We can use the \citet{bruzualcharlot03} models to find over what timescale a galaxy will transform from blue to red once star formation ceases. We plot ($B-R$) and ($V-I$) colours versus time in Fig.~\ref{bruzualcharlot}. The ($B-R$) colours are included for reference as this colour will evolve faster than the ($V-I$) colour. It can be seen that a blue stellar population evolves rapidly to a red one, with a galaxy transforming from blue to red over a timescale of $\sim$0.5 Gyr. If ram pressure stripping is the dominant mechanism driving the transformation from blue to red galaxies, then the complete transformation from a blue, star forming dwarf to a red, passive dwarf will take 0.5 - 1 Gyr.

However, to completely transform from a late to early type, internal sub-structure must be removed from the galaxy. The dwarfs in the outer regions contain asymmetries and clumpiness in their light distribution, whereas those in the core do not, so if the two populations of dwarfs have the same origin, clearly this structure must somehow be removed. We can estimate over what timescale star clusters within the dwarf will dissipate, removing substructure caused by star formation. Stars typically form in clusters of $M \sim 10^{3}$ M$_{\odot}$, and these will disperse and dissolve over time. Such clusters in dense environments such as the center of the MW will dissolve rapidly, on a timescale of $\sim10$ Myr \citep{kim00}. However, dwarf galaxies typically have stellar densities much lower than in more massive galaxies, and therefore the timescale for star clusters to evaporate will be longer. Open clusters in the outer regions of the MW are found to have ages between 0.5-6 Gyr (e.g. \citealt{hasegawa08}). Also, substructure such as spiral arms and bars has been found for two dEs in the Fornax Cluster \citep{sven03}, with ages $>2$ Gyr determined for these dwarfs via spectroscopy \citep{michielsen07}. It is likely therefore that structure will remain in a dwarf galaxy that has ceased star formation for at least a Gyr, though detailed simulations will be required to obtain a more precise life-time. Therefore, the morphological transformation from late to early type galaxy will take longer than the colour transformation.

Other processes will also play a role in the transformation from late-to-early type galaxy to a lesser extent. For example galaxy harassment \citep{moore} caused by encounters with larger galaxies and the cluster's gravitational potential, will transform a disky late-type galaxy into a dE or dSph over a timescale of 3-4 Gyr. Such an encounter will cause the internal kinetic energy of the infalling galaxy to increase, modifying its energy and mass profile. To counteract this and regain equilibrium, the size of the galaxy will increase and it loses its most energetic stars. Over time, the galaxies density decreases, resulting in the morphological transformation of a spiral to a dE over several orbits ($\sim2$-$4$ Gyr).

The number counts of faint dwarfs (M$_{V} < -13$) suggest that in the dense cluster core, environmental influences become very important, with only the most dark matter dominated dwarfs with sufficient mass to survive the cluster tidal forces able to survive in this region. These forces are lower in the cluster outskirts, resulting in higher number counts compared to the core region.

The result that the lowest mass dwarfs are preferentially found in less dense regions has been observed for other clusters. \citet{Pra} find that the luminosity function in Abell 2218 changes from $\alpha = -1.23$ in the cluster core to $\alpha = -1.49$ outside the radius $r = 300$ kpc. Likewise \citet{sanchezjanssen05} find a similar result for the Hercules Cluster (A2151), with the radial LF showing a trend of increasing faint-end slope as cluster-centric radius increases. This result is likely due to tidal processes, which are more effective at smaller cluster-centric radii.

\section{Conclusions}

We identify from $HST$ ACS imaging a total of 11 dwarf ellipticals in the outskirts of the Perseus Cluster, all of which are newly discovered. We compare the colours, structures and morphologies of these newly discovered dwarfs in the cluster outskirts to those in the core region, for which we have $HST$ ACS imaging. To quantify their morphologies, we compute their concentration, asymmetry and clumpiness (CAS) parameters and compare these results to those obtained in \citet{me09} for the dwarfs in the core. The dwarfs in the cluster outskirts have, on average, more disturbed structures than those in the core, with $<$A$>$ $=0.13\pm0.09$ and $<$S$>$ $=0.18\pm0.08$, compared to $<$A$>$ $=0.02\pm0.04$, $<$S$>$ $=0.01\pm0.07$ for those in the cluster core. 

The fact that the dwarfs in the cluster outskirts are morphologically disturbed whereas those in the core are not suggests that those in the outskirts are part of an infalling population of galaxies transitioning from dIrr to dE, and not part of an original population. We show that these galaxies lie on the same red sequence as those in the cluster core, showing they have ceased star formation. Further evidence for this lack of star formation comes from our $GALEX$ imaging, whereby none of the dwarfs in the outer regions are detected in our NUV imaging, indicating they have SFRs of less than 0.001 M$_{\odot}$yr$^{-1}$. From this, and the fact that these dwarfs have more disturbed morphologies than those in the cluster core, we infer and show that the colours of these galaxies transform over a quicker timescale than their morphologies.  By comparing the colours of these dwarfs to \citet{bruzualcharlot03} models of colour versus the time since star formation ceased, we find that the colours of these galaxies can transform rapidly over a timescale of $\sim$500 Myr, whereas substructure likely remains in the galaxies for several Gyr after star formation ceases.

Finally, to examine if local galaxy density has an effect on the faint end slope of the luminosity function, we investigate how the number counts in the dense core of Perseus compare to those in the outskirts. The number counts for faint galaxies ($M_{V} < -12$) are higher in the cluster outskirts than in the core, suggesting that the faintest galaxies are unable to survive in the very dense parts of the cluster unless they are sufficiently dark matter dominated to prevent their disruption by the cluster potential. This provides further evidence for the existence of a dwarf galaxy density relation, whereby a steeper faint-end slope of the luminosity function is found in the lower density environments.

\section*{Acknowledgments}

SJP acknowledges the support of a STFC studentship. EVH acknowledges support from an ASI-COFIS grant. RWO acknowledges support from the GALEX Guest Investigator program administered by NASA--Goddard Space Flight Center. Support for the US participants in programs GO-10201 and GO-10789 was provided by NASA through a grant from the Space Telescope Science Institute, which is operated by the Association of Universities for Research in Astronomy, Inc., under NASA contract NAS 5-26555. This research has made use of the NASA/IPAC Extragalactic Database (NED) which is operated by the Jet Propulsion Laboratory, California Institute of Technology, under contract with the National Aeronautics and Space Administration. We would like to thank Ruth Gr\"{u}tzbauch and Asa Bluck for useful discussions.


\begin{thebibliography}{99}
\bibitem[\protect\citeauthoryear{Aguerri \& Gonz\'{a}lez-Garc\'{i}a}{2009}]{aguerri09} Aguerri J.A.L., Gonz\'{a}lez-Garc\'{i}a, A.C., 2009, A\&A, 494, 891
\bibitem[\protect\citeauthoryear{Beasley et al.}{2009}]{beasley09} Beasley M.A., Cenarro A.J., Strader J., Brodie J.P., 2009, AJ, 137, 5146
\bibitem[\protect\citeauthoryear{Bekki \& Couch}{2003}]{bekki} Bekki K., Couch W.J., 2003, ApJ, 596, L13
\bibitem[\protect\citeauthoryear{Bertin \& Arnouts}{1996}]{SExtractor} Bertin E., Arnouts S., 1996, A\&AS, 117, 393
\bibitem[\protect\citeauthoryear{Binggeli, Tammann \& Sandage}{Binggeli et al.}{1987}]{binggeli87} Binggeli B., Tammann G.A., Sandage A., 1987, AJ, 94, 251
\bibitem[\protect\citeauthoryear{Boselli et al.}{2005}]{boselli} Boselli A. et al., 2005, ApJ, 629, L29
\bibitem[\protect\citeauthoryear{Bouchard et al.}{2007}]{bouchard07} Bouchard A., Jerjen H., Da Costa G.S., Ott J., 2007, AJ, 133, 261
\bibitem[\protect\citeauthoryear{Bressan, Chiosi \& Fagotto}{1994}]{bressan94} Bressan A., Chiosi C., Fagotto G., 1994, ApJS, 94, 63
\bibitem[\protect\citeauthoryear{Bruzual \& Charlot}{2003}]{bruzualcharlot03} Bruzual G., Charlot S., 2003, MNRAS, 344, 1000
\bibitem[\protect\citeauthoryear{Conselice, Bershady \& Jangren}{Conselice et al.}{2000}]{Conselice00} Conselice C.J., Bershady M.A., Jangren A., 2000, ApJ, 529, 886
\bibitem[\protect\citeauthoryear{Conselice}{2003}]{cas} Conselice C.J., 2003, ApJS, 147, 1
\bibitem[\protect\citeauthoryear{Conselice, Gallagher \& Wyse}{Conselice et al.}{2001}]{cgw01} Conselice C.J., Gallagher J.S. III, Wyse R.F.G., 2001, ApJ, 559, 791
\bibitem[\protect\citeauthoryear{Conselice, Gallagher \& Wyse}{Conselice et al.}{2002}]{cgw02} Conselice C.J., Gallagher J.S. III, Wyse R.F.G., 2002, AJ, 123, 2246
\bibitem[\protect\citeauthoryear{Conselice, Gallagher \& Wyse}{Conselice et al.}{2003a}]{cgw03} Conselice C.J., Gallagher J.S. III, Wyse R.F.G., 2003, AJ, 125, 66
\bibitem[\protect\citeauthoryear{Conselice, Gallagher \& Wyse}{Conselice et al.}{2003b}]{cgw04} Conselice C.J., Gallagher J.S. III, Wyse R.F.G., 2003, ApJ, 591, 167
\bibitem[\protect\citeauthoryear{Cortese et al.}{2006}]{cortese} Cortese L., Gavazzi G., Boselli A., Franzetti P., Kennicutt R.C., O'Neil K., Sakai, S., 2006, A\&A, 453, 847
\bibitem[\protect\citeauthoryear{C\^{o}t\'{e} et al.}{2006}]{cote06} C\^{o}t\'{e} P., et al., 2006, ApJS, 165, 57
\bibitem[\protect\citeauthoryear{De Rijcke \& Dejonghe}{2002}]{sven02} De Rijcke S., Dejonghe H., 2002, MNRAS, 329, 829
\bibitem[\protect\citeauthoryear{De Rijcke et al.}{2003}]{sven03} De Rijcke S., Dejonghe H., Zeilinger W.W., Hau G.K.T., 2003, A\&A, 400, 119
\bibitem[\protect\citeauthoryear{De Rijcke et al.}{2009}]{sven09} De Rijcke, S., Penny S.J., Conselice C.J., Valcke S., Held E.V., 2009, MNRAS, 393, 798
\bibitem[\protect\citeauthoryear{Dressler}{1980}]{dressler80} Dressler A., 1980, ApJ, 236, 351
\bibitem[\protect\citeauthoryear{Ferguson \& Sandage}{1989}]{ferguson89} Ferguson H.C., Sandage A., 1989, ApJ, 346, L53
\bibitem[\protect\citeauthoryear{Gallagher \& Hunter}{1989}]{gallagherhunter} Gallagher J.S. III, Hunter D.A., 1989, AJ, 98, 806
\bibitem[\protect\citeauthoryear{Grant, Kuipers \& Phillipps}{Grant \& Phillipps}{2005}]{grant05} Grant N.I., Kuipers J.A., Phillipps S., 2005, MNRAS, 363, 1019
\bibitem[\protect\citeauthoryear{Grebel, Gallagher \& Harbeck}{2003}]{grebel03} Grebel E.K., Gallagher J.S. III, Harbeck D., 2003, AJ, 125, 1926
\bibitem[\protect\citeauthoryear{Hasegawa, Sakamoto \& Malasan}{Hasegawa et al.}{2008}]{hasegawa08} Hasegawa T., Sakamoto T., Malasan H.L., 2008, PASJ, 60, 1267
\bibitem[\protect\citeauthoryear{Holtzman et al.}{1995}]{wfpc2cal} Holtzman A., Burrows C.J., Casertano S., Hester, J.J., Trauger, J.T., Watson, A.M., Worthey G., 1995, PASP, 107, 1065
\bibitem[\protect\citeauthoryear{Jerjen, Kalnajs \& Binggeli}{Jerjen et al.}{2000}]{jerjen00} Jerjen H., Kalnajs A., Binggeli B., 2000, A\&A, 358, 2000
\bibitem[\protect\citeauthoryear{Kennicutt}{1998}]{kennicutt98} Kennicutt R.C., 1998, ARAA, 36, 189
\bibitem[\protect\citeauthoryear{Kim et al.}{2000}]{kim00}Kim S. S., Figer D. F., Lee H. M.,  Morris M., 2000, ApJ, 545, 301
\bibitem[\protect\citeauthoryear{Kormendy}{1985}]{kormendy85} Kormendy J., 1985, ApJ, 295, 73
\bibitem[\protect\citeauthoryear{Kormendy et al.}{2009}]{kormendy09} Kormendy J., Fisher D. B., Cornell M. E., Bender R., 2009, ApJS, 182, 216
\bibitem[\protect\citeauthoryear{Kron}{1995}]{Kron95} Kron R.G., 1995, The Deep Universe: Saas-Fee Advanced Course 23. Springer, New York
\bibitem[\protect\citeauthoryear{Leauthaud et al.}{2007}]{cosmos} Leauthaud A., et al., 2007, ApJS, 172, 219
\bibitem[\protect\citeauthoryear{Lisker et al.}{2006}]{Lisker06} Lisker T., Grebel E.K., Binggeli B., 2006, AJ, 132, 497
\bibitem[\protect\citeauthoryear{Lisker \& Fuchs}{2009}]{lisker09a} Lisker T., Fuchs B., 2009, A\&A, 501, 429
\bibitem[\protect\citeauthoryear{Lisker et al.}{2009}]{lisker09} Lisker T., et al., 2009, ApJ, 706, L124
\bibitem[\protect\citeauthoryear{Lotz, Miller \& Ferguson}{Lotz et al.}{2005}]{lotz05} Lotz J.M., Miller B.W., Ferguson H.C., 2004, ApJ, 613, 262
\bibitem[\protect\citeauthoryear{Michielsen, De Rijcke \& Dejonghe}{Michielsen et al.}{2004}] {michielsen04} Michielsen D., De Rijcke S., Dejonghe H., 2004, Astron. Nachr. Supp., 325, 122 
\bibitem[\protect\citeauthoryear{Michielsen et al.}{2007}]{michielsen07} Michielsen D., Koleva M., Prugniel P., Zeilinger W.W., De Rijcke S., Dejonghe H., Pasquali A.,  Ferreras I., Debattista V.P., 2007, ApJ, 670, 10
\bibitem[\protect\citeauthoryear{Michielsen et al.}{2008}]{michielsen08} Michielsen D., et al., 2008, MNRAS, 385, 1374
\bibitem[\protect\citeauthoryear{Mieske et al.}{2007}]{mieske07} Mieske S., Hilker M., Infante L., Mendes de Oliveira C., 2007, A\&A, 463, 503
\bibitem[\protect\citeauthoryear{Misgeld, Mieske \& Hilker }{Misgeld et al.}{2007}]{misgeld08} Misgeld I., Mieske S., Hilker M., 2008, A\&A, 486, 697 
\bibitem[\protect\citeauthoryear{Moore et al.}{1996}]{moore96} Moore B., Katz N., Lake G., Dressler A., Oemler A., Nature, 1996, 379, 613
\bibitem[\protect\citeauthoryear{Moore, Lake \& Katz}{Moore et al.}{1998}]{moore} Moore B., Lake G., Katz N., 1998, ApJ, 495, 139
\bibitem[\protect\citeauthoryear{Mori \& Burkert}{2000}]{mori00} Mori M., Burkert A., 2000, ApJ, 538, 559
\bibitem[\protect\citeauthoryear{Penny \& Conselice}{2008}]{me08} Penny S.J., Conselice C.J., 2008, MNRAS, 383, 247
\bibitem[\protect\citeauthoryear{Penny et al.}{2009}]{me09} Penny S.J., Conselice C.J., De Rijcke S., Held E.V., 2009, MNRAS, 393, 1054
\bibitem[\protect\citeauthoryear{Petrosian}{1976}]{Pet76} Petrosian V., 1976, ApJ, 209, L1
\bibitem[\protect\citeauthoryear{Phillips et al.}{1998a}]{phillips98a} Phillipps S., Parker Q.A., Schwartzenberg J.M., Jones J.B., 1998, ApJ, 493, L59
\bibitem[\protect\citeauthoryear{Phillips et al.}{1998b}]{phillips98} Phillips S., Driver S.P., Couch W.J., Smith R.M., 1998, ApJ, 498, L119
\bibitem[\protect\citeauthoryear{Poggianti et al.}{2001}]{poggianti01} Poggianti B.M. et al, 2001, ApJ, 562, 689
\bibitem[\protect\citeauthoryear{Pracy et al.}{2004}]{Pra} Pracy M. B., De Propris R., Driver S. P., Couch W. J., Nulsen P. E. J., 2004, MNRAS, 352, 1135
\bibitem[\protect\citeauthoryear{Reaves}{1962}]{reaves62} Reaves G., 1962, PASP, 74, 392
\bibitem[\protect\citeauthoryear{S\'{a}nchez-Janssen et al.}{2005}]{sanchezjanssen05} S\'{a}nchez-Janssen R., Iglesias-P\'{a}ramo J., Mu\~{n}oz-Tu\~{n}\'{o}n C., Aguerri J. A. L., V\'{i}lchez J. M., 2005, A\&A, 434, 521
\bibitem[\protect\citeauthoryear{Sandage}{1972}]{sandage72} Sandage A., 1972, ApJ, 176, 21
\bibitem[\protect\citeauthoryear{Sandage \& Hoffman}{1991}]{sandage91} Sandage A., Hoffman G.L., 1991, ApJ, 379, 45 
\bibitem[\protect\citeauthoryear{Schlegel et al.}{1998}]{schlegel} Schlegel D.J., Finkbeiner D.P., Davis M., 1998, ApJ, 500, 525
\bibitem[\protect\citeauthoryear{Sirianni et al.}{2005}]{sirianni05} Sirianni M., et al., 2005, PASP, 117, 1049
\bibitem[\protect\citeauthoryear{Smith Castelli et al.}{2008}]{smithcastelli} Smith Castelli A.V.,  Bassino L.P., Richtler T., Cellone S.A., Aruta C., Infante L., 2008, MNRAS, 386, 2311
\bibitem[\protect\citeauthoryear{Struble \& Rood}{1999}]{Stublerood99} Struble M.F., Rood H.J., 1999, ApJS, 125, 36
\bibitem[\protect\citeauthoryear{Thompson \& Gregory}{1993}]{thompson} Thompson L.A., Gregory S.A., 1993, AJ, 106, 2197
\bibitem[\protect\citeauthoryear{Trentham \& Tully}{2009}]{trent09} Trentham T., Tully R.B., 2009, MNRAS, 398, 722
\bibitem[\protect\citeauthoryear{van den Bergh}{1994}]{vdb94} van den Bergh S., 1994, AJ, 107, 1328
\end{thebibliography}
\end{document}